\documentclass[aps,prd]{revtex4}
\usepackage{subfigure,graphics} 
\usepackage{flafter} 
\begin{document} 

\title{
On the consistency of warm inflation} 
\author{Ian G. Moss}
\email{ian.moss@ncl.ac.uk}
\author{Chun Xiong}
\affiliation{School of Mathematics and Statistics, Newcastle University,
Newcastle Upon Tyne, NE1 7RU, UK}

\date{\today}

%%%%%%%%%%%%%%%%%%%%%%%%%%%%%%%%%%%%%%%%%%%%%%

\begin{abstract}
Conditions are obtained for the existence of a warm inflationary attractor in the system of
equations describing an inflaton coupled to radiation. These conditions restrict the temperature
dependence of the dissipative terms and the size of thermal corrections to the inflaton potential,
as well as the gradient of the inflaton potential. When these conditions are met, the evolution
approaches a slow-roll limit and only curvature fluctuations survive on super-horizon scales.
Formulae are given for the spectral indices of the density perturbations and the tensor/scalar
density perturbation amplitude ratio in warm inflation.
\end{abstract}
\pacs{PACS number(s): }

\maketitle
%%%%%%%%%%%%%%%%%%%%%%%%%%%%%%%%%%%%%%%%%%%
\section{introduction}

Inflationary models \cite{guth81,linde82,albrecht82} have proved very sucessful in explaining many
of the large scale features of the universe (see e.g. \cite{Liddle:2000cg}). An essential feature
of these inflationary models is their
stability, meaning in particular that inflation solutions are attractors in the solution space of
the relevant cosmological equations (see e.g. \cite{Salopek:1990jq,Liddle:1994dx}),
at least up until inflation ends and the universe becomes radiation dominated. Without this
feature, inflation might never have begun, and certainly would not have lasted long enough to
affect the large scale structure of the universe.

Warm inflation is an alternative inflationary scenario in which a small but significant amount of
radiation survives during the inflationary era due to continuous particle production
\cite{Moss85,bererafang95,berera95}. The coupling
between radiation and the inflaton field leads to thermal dissipation and fluctuations in the time
evolution of the inflaton field. The stability of the inflationary solutions in warm inflationary
models has only been addressed in a limited form previously \cite{deOliveira:1997jt}, and we will
present the full stability analysis here. We shall examine conditions under which warm inflation is
an attractor and give conditions for a prolonged
period of warm inflation.

We shall show that the stability of warm inflation can be related to conditions on two parameters
describing the temperature dependence of terms in the inflaton equation of motion which where not
taken into account in the earlier stability analysis \cite{deOliveira:1997jt}. 
The first condition says that if the dissipation term
in the equation of motion falls off too rapidly at low temperature then the temperature is driven to
zero, and we fall into the conventional inflationarty scenario. The second condition limits the
temperature dependence of the inflaton potential. In many models
large dissipation implies large thermal corrections to the potential which prevent warm inflation
taking place. This argument is essentially the one first presented in Ref. \cite{yokoyama99}.
Nowadays, we know that there are models where thermal corrections to the inflaton potential are
suppressed by supersymmetry,  and these models may allow warm inflation as an attractor 
\cite{BasteroGil:2006vr,BuenoSanchez:2008nc,review}.

The duration of the period of inflation is related to a set of slow-roll paramaters which where
introduced in \cite{Hall:2003zp}. We shall re-derive the slow-roll conditions for warm inflation as
part of the stability analysis. A well understood feature of the slow-roll conditons for warm
inflation is that they can be less restrictive than the slow-roll condtions for conventional
inflation. 

We stress that we are concerned here with the self-consistency of the warm inflationary scenario for
given equations of motion. We shall not address how the equation of motion for the inflaton field
is obtained from non-equilibrium thermal field theory. A discussion of the derivation of the
equations of motion can be found in a recent review \cite{review}. However, we would like to point
out that some of the critisims of warm inflation have been based on models which 
do not satisfy the fundamental stability conditions derived here, and are therefore not inconsistent
with the validity of warm inflation in general \cite{Aarts:2007ye}.

The stability of warm inflation has consequences for the origin and evolution of cosmological
density fluctuations \cite{Hall:2003zp}. In warm inflation, density fluctuations originate from
thermal fluctuations \cite{bererafang95,berera00}. In particular, the fact that the inflationary
solution depends on only one parameter means that only one perturbation mode,
the curvature perturbation, survives on super-horizon scales despite the fact that there are
entropy perturbations present on sub-horizon scales. We shall give formulae for the spectral indices
of the scalar and tensor modes and say a little about the tensor/scalar ratio.

\section{basic equations}
\label{be}

We start with a flat, homogeneous universe with expansion rate $H$. The matter content consists of a
homogeneous inflaton field $\phi$ and thermal radiation of temperature $T$. We restrict attention to
the warm inflationary regime where $T>H$, with the radiation close to thermal equilibrium. The
evolution of the inflaton is governed by a potential $V(\phi,T)$ and a damping coefficient
$\Gamma(\phi,T)$, such
that the inflaton field satisfies the basic equation
\begin{equation}
\ddot\phi+(3H+\Gamma)\dot\phi+V_{,\phi}=0,\label{phieq}
\end{equation}
where a comma after a function denotes a derivative. The expansion rate is related to the energy
density $\rho$ by the Friedman equation
\begin{equation}
3H^2=8\pi G\rho,\label{heq}
\end{equation}
where $G$ is Newton's constant.

It is important to realise that the potential appearing in the inflaton equation is the free energy
density, rather than the potential energy density. The potential energy density is given by the
thermodynamic relation $V+Ts$, where $s$ is the entropy density
\begin{equation}
s=-V_{,T}.\label{sdef}
\end{equation}
The total energy density, including the inflaton's kinetic energy density, is
therefore
\begin{equation}
\rho=\frac12\dot\phi^2+V+Ts.\label{rhoeq}
\end{equation}
This includes the contributions of the scalar field and the radiation. It is not always possible to
separate the scalar and radiation components of the energy density in an unambiguous way.

The final equation is the one which governs the transfer of energy from the inflaton to the
radiation field. This can easily be derived from the stress-energy tensor
$T_{\mu\nu}$ \cite{Hall:2003zp},
\begin{equation}
T_{\mu\nu}=Ts\,u_\mu u_\nu-V g_{\mu\nu}+
\nabla_\mu\phi\nabla_\nu\phi-\frac12(\nabla\phi)^2 g_{\mu\nu},
\end{equation}
where $u_\mu$ is the radiation fluid $4-$velocity vector, $g_{\mu\nu}$ is the metric and
$\nabla_\nu$ is the spacetime
derivative operator. Conservation of the stress-energy gives
\begin{equation}
T(\dot s+3Hs)=\Gamma\dot\phi^2.\label{seq}
\end{equation}
This version of the second law of thermodynamics shows clearly how the friction converts the
inflaton's energy into heat. We now have a complete set of evolution equations which can be solved
for  $\phi$ and $s$ given a set of initial conditions.

Inflation is associated with a slow-roll approximation which consists of dropping the leading
derivative term in each equation. The slow-roll equations are therefore
\begin{eqnarray}
\dot\phi&=&{-V_{,\phi}\over 3H(1+Q)}\label{sr1}\\
Ts&=&Q\dot\phi^2\label{sr2}\\
H^2&=&{8\pi G V\over 3}\label{sr3}
\end{eqnarray}
where the strength of the dissipation is quantified by the parameter $Q$,
\begin{equation}
Q={\Gamma\over 3H}.
\end{equation}
Note that, since Eq. (\ref{sr1}) is first order in time derivatives, any solution to the slow-roll
equations has just one constant of integration.

The validity of the slow-roll approximation will depend on the size of a set of slow-roll
parameters \cite{Hall:2003zp}. We use the following set of `small' parameters:
\begin{eqnarray}
\epsilon={1\over 16\pi G}\left({V_{,\phi}\over V}\right)^2,\quad
\eta={1\over 8\pi G}{V_{,\phi\phi}\over V},\quad
\beta={1\over 8\pi G}{V_{,\phi}\Gamma_{,\phi}\over V\Gamma}.
\end{eqnarray}
An additional pair of parameters describe the temperature dependence,
\begin{equation} 
b={TV_{,\phi T}\over V_{,\phi}},\quad
c={T\Gamma_{,T}\over \Gamma}\label{srp}.
\end{equation}
We use $b$ in place of the parameter $\delta$ defined in ref
\cite{Hall:2003zp}.

The parameter $b$ is very important in the theory of warm inflation. It measures the size of
contributions to the potential from thermal quantum fields. In ordinary circumstances, we would
expect $b$ to be order 1. Consider, for example, an inflaton of mass $m_\phi$ coupled to a set of
$g_*$ light scalar fields and temperature $T>H>m_\phi$. The effective potential,
\begin{equation}
V(\phi,T)=-{\pi^2\over 90}g_*T^4-{1\over 12}m_\phi^2T^2
+v(\phi),\label{corpot}
\end{equation}
where $v(\phi)$ is the effective potential at $T=0$. The parameter $b$ for the potential
(\ref{corpot}) is approximately 2. Values of $b$ can be much smaller in supersymmetric theories,
where there is a cancellation of leading order thermal corrections when the temperature
$T<\Lambda_S$, where $\Lambda_S$ is the supersymmetry breaking scale \cite{Hall:2004zr}. We shall
find limits on $b$, and conclude that  warm inflation only takes place when there is a mechanism,
like supersymmetry,
which reduces the size of the thermal corrections to the potential \cite{berera02}.

\section{stability analysis}

We shall consider the consistency of the slow-roll approximation by performing a linear stability
analysis to determine the conditions which are sufficient for the system to remain close to the
slow-roll solution for many Hubble times. It may be worthwhile considering first what happens in the
alternative cold inflationary scenario (see e.g. \cite{Liddle:2000cg}).
The slow-roll equation in this case is first order in time derivatives, and the general solution
has the form $\phi=f(t-t_0)$, where $t_0$ is an arbitrary constant. There always exists a
homogeneous perturbation which is equivalent to changing the value of $t_0$, and is hugely
important for the existence of density perturbations. Another mode decays on the Hubble timescale.
It proves convenient to exclude the time-translation mode
by using the value of the field as the time coordinate, which is possible if the inflaton field has
non-vanishing time derivative. We shall follow the same procedure for the analysis or warm
inflation. 

With the inflaton as independent variable we can rewrite the system of equations in first order
form,
\begin{equation}
x'=F(x).
\end{equation}
Prime denotes derivatives with respect to $\phi$ and
\begin{equation}
x=\pmatrix{u\cr s}
\end{equation}
where $u=\dot\phi$. Eqs. (\ref{phieq}) and (\ref{seq}) become
\begin{eqnarray}
u'&=&-3H-\Gamma-V_{,\phi}u^{-1},\\
s'&=&-3Hsu^{-1}+T^{-1}\Gamma u,
\end{eqnarray}
with the temperature determined implicitly by Eq. (\ref{sdef}) and $H$ given by Eq. (\ref{heq}). We
take a background $\bar x$ which satisfies the slow-roll equations (\ref{sr1}-\ref{sr3}) and then
the linearised perturbations satisfy
\begin{equation}
\delta x'=M(\bar x)\delta x-{\bar x}'.\label{dxeq}
\end{equation}
where $M$ is the matrix of first derivatives of $F$ evaluated at the slow-roll solution.

Consider stability first of all. We can express all the components of the matrix $M$ in terms of the
slow-roll parameters (\ref{srp}). For
example,
\begin{equation}
\Gamma_{,s}=\Gamma_{,T}T_{,s}={c\Gamma\over T}T_{,s}=
{c\over 3}{\Gamma\over s}=cQ{H\over s}.
\end{equation}
where we have used $T_{,s}=(s_{,T})^{-1}=T/3s$. Following a similar procedure for all of the first
derivatives gives a final expression for $M$,
\begin{equation}
M=\pmatrix{
A&B\cr C&D\cr
}
\end{equation}
where
\begin{eqnarray}
A&=&{H\over u}\left\{
-3(1+Q)-{\epsilon\over (1+Q)^2}\right\}\\
B&=&{H\over s}\left\{
-cQ-{Q\over(1+Q)^2}\epsilon+(1+Q)b\right\}\\
C&=&{Hs\over u^2}\left\{
6-{\epsilon\over (1+Q)^2}\right\}\\
D&=&{H\over u}\left\{
c-4-{Q\epsilon\over (1+Q)^2}\right\}\\
\end{eqnarray}
We require the determinant,
\begin{equation}
\det M={H^2\over u^2}\left\{
12(1+Q)+3(Q-1)c-6(1+Q)b
+\left({3Q^2+9Q+4\over (1+Q)^2}-{c\over 1+Q}\right)\epsilon
+{b\epsilon\over 1+Q},
\right\}
\end{equation}
and the trace,
\begin{equation}
{\rm tr}\,M={H\over u}\left\{
c-4-3(1+Q)-{\epsilon\over 1+Q}
\right\}.
\end{equation}
In the cold inflationary case, where $Q=b=c=s=0$, the decaying modes have the approximate form
$\delta u\propto \exp(-3N)$ and $\delta s\propto \exp(-4N)$, where $N$ is the number of e-folds of
inflation \cite{Salopek:1990jq}.

{\em Sufficient} conditions for stability of the warm inflationary solution are that the matrix $M$
varies slowly and 
\begin{equation}
|c|\le 4-2b,\qquad b\ge 0.
\end{equation}
Evidently, the trace is negative and the determinant is positive if these conditions hold. The
linear equation therefore has two negative eigenvalues and the both eigenmodes decay. 
The slow variation of $M$, which allows us to diagonalise the linear system, follows if the forcing
term in Eq. (\ref{dxeq}) is small, and so we turn to this term next.

The forcing term in Eq. (\ref{dxeq}) depends on $\bar x'$. This term is present because the
background chosen is not an exact solution to the full set of equations. The slow-roll approximation
can only be valid when $\bar x'$ is small. If we work with time
derivatives, the magnitude of $\bar x'$ depends on the the dimensionless quantities 
$\dot u/(Hu)$ and $\dot s/(Hs)$, and small values represent slow variation on the timescale of the
Hubble time.

We only quote the leading terms in $\epsilon$. Starting from the slow-roll Eq. (\ref{sr3}) and
taking the time derivative gives
\begin{equation}
{\dot H\over H^2}=-{1\over 1+Q}\epsilon.\label{hdot}
\end{equation}
By combining the other slow -roll equations (\ref{sr1}) and (\ref{sr2}) we eventually arive at 
\begin{eqnarray}
{\dot u\over Hu}&=&
{1\over \Delta}\left\{
-3c(1+Q)b
-{c(1+Q)-4\over 1+Q}\epsilon+(c-4)\eta+{4Q\over 1+Q}\beta
\right\}\label{udot}\\
{\dot s\over Hs}&=&
{1\over \Delta}\left\{
+{3(cQ+Q+1-c)(1+Q)\over Q}b
+{3Q+9\over 1+Q}\epsilon-6\eta+{3(Q-1)\over 1+Q}\beta
\right\}\label{sdot}
\end{eqnarray}
where $\Delta=4(1+Q)+(Q-1)c$. The slow-roll approximation requires $\dot u<<Hu$ and 
$\dot s<<Hs$, and sufficient conditions for this are
\begin{equation}
\epsilon,\ |\beta|\ ,|\eta|\ <<1+Q;\ 0<b<<{Q\over 1+Q},\ |c|<4.
\end{equation}
The conditions on $\epsilon$ and $\eta$ agree with a previous stabilty analysis
\cite{deOliveira:1997jt}. These are weaker than the corresponding conditions for cold
inflation, and this fact is a well know feature of warm inflation.

The physical interpretation of
the condition $c<4$ is evident from Eq. (\ref{seq}), that radiation must be produced at a rate
($\Gamma\propto T^c$) exceeding the rate at which radiation is removed by the expansion of the
universe ($Ts \propto T^4$). The condition on $b$ can only be met if
there is a mechanism for suppressing thermal corrections to
the potential because, as we mentioned at the end of Sect \ref{be}, high temperature thermal
corrections would
otherwise lead to $b\approx 2$. Models which include a mechanism for suppressing thermal corrections
can be found, for example, in
\cite{berera02,BasteroGil:2006vr,BuenoSanchez:2008nc}.

\section{density fluctuations}

The results which have been obtained as part of the stability analysis are also helpful for
analysing various features of the density fluctuation spectrum. We therefore take the
opportunity, whilst the results are to hand, of giving some formulae which might be useful for
observational tests of warm inflation.

The origin of density fluctuations in warm inflationary scenarios is due to thermal fluctuations in
the radiation. These are coupled to the inflaton as a consequence of the friction term in the
inflaton equation of motion, and their amplitude is fixed by a fluctuation-dissipation theorem.
This means that both entropy and curvature perturbations must be present. However, on length scales
larger than the horizon, we know from the stability argument that the coupled inflaton plus
radiation system approaches the slow-roll solution which has only one free parameter. Consequently,
on large scales only the pure curvature perturbation survives. This has been confirmed in
particular models by solving the full set of density fluctuation equations numerically
\cite{Hall:2003zp}.

Even though the entropy perturbations decay on large scales, they can sometimes leave behind an
impression on the curvature fluctuations. If the friction term depends on temperature, then the
entropy and curvature fluctuations on sub-horizon scales become coupled. The situation is similar to
the sympathetic oscillations of a double pendulum \cite{som}. When the curvature fluctuations
`freeze-out', the amplitude of the sympathetic oscillation may be anywhere between zero and its
maximum value, leading to oscillations in the wave-number dependence of the curvature fluctuation
spectrum. The amplitude given below therefore has only limited use when $b$ and $c$ are non-zero
and refers only to the envelope of these oscillations.

The thermal fluctuations produce a power spectrum of scalar density fluctuations of the form 
\cite{Hall:2003zp,Moss:2007cv},
\begin{equation}
{\cal P}_S={\sqrt{\pi}\over 2}{H^3T\over u^2}(1+Q)^{1/2}.\label{ps}
\end{equation}
The spectral index $n_s$ is defined by
\begin{equation}
n_S-1={\partial \ln{\cal P}_S\over \partial \ln k},
\end{equation}
evaluated when the amplitudes `freeze out'. To leading order in the slow-roll parameters we can take
the freeze-out time to be the horizon crossing time when $k=aH$, and then
\begin{equation}
n_S-1={\dot{\cal P}_S\over H{\cal P}_S}.
\end{equation}
We can use eqs. (\ref{hdot}-\ref{sdot}) to obtain
\begin{equation}
n_S-1={1\over\Delta}
\left\{
-{3(2Q+2+5Qc)(1+Q)\over Q}b-
{9Q+17-5c\over 1+Q}\epsilon
-{9Q+1\over 1+Q}\beta
-{3Qc-6-6Q+2c\over 1+Q}\eta
\right\}
\end{equation}
 If we consider $b=c=0$, then important limits include the strong regime of warm inflation, $Q>>1$,
\begin{equation}
n_S-1=-{9\over 4Q}\epsilon-{9\over 4Q}\beta+{3\over 2Q}\eta.
\end{equation}
This agrees with a partial result in \cite{taylor00} and the full result in
\cite{Hall:2003zp}. In the weak regime of warm inflation, $Q<<1$, thermal fluctuations lead to the
spectral index 
\begin{equation}
n_S-1=-{17\over 4}\epsilon-{1\over 4}\beta+{3\over 2}\eta.
\end{equation}
Previous results for the weak regime, though expressed in a less useful form, can be found in Refs
\cite{BasteroGil:2004tg,Hall:2007qw}. Finally, the case $Q<<1$ and $c=3$ is important because it
corresponds to a class of warm
inflationary models where the friction coefficient $\Gamma$ has been calculated \cite{Moss:2006gt},
\begin{equation}
n_S-1=-2\epsilon-\beta.
\end{equation}

The tensor modes have the same amplitude as they do in the cold inflationary models,
\footnote{
Our power spectrum convention for $\delta_k$ is 
$\langle\delta_k\delta_{k'}\rangle=(2\pi)^3k^{-3}{\cal P}_\delta(k)\delta({\bf k}_1-{\bf k}_2)$.
}
\begin{equation}
{\cal P}_T=8\pi G H^2.
\end{equation}
The spectrum for the tensor modes is simply
\begin{equation}
n_T-1=-{2\over 1+Q}\epsilon.
\end{equation}
Unlike in the cold inflationary scenario, the tensor-scalar amplitude ratio cannot be expressed in
terms of slow-roll parameters. Instead \cite{taylor00},
\begin{equation}
{{\cal P}_T\over {\cal P}_S}=
{2\epsilon\over (1+Q)^3}{H\over T}.
\end{equation}
Since $T>H$ for warm inflation, the tensor-scalar ratio is likely to be smaller than 
$1-n_T$. Tensor modes are strongly suppressed relative to the scalar modes on the strong regime of
warm inflation $Q>>1$, but they could be significant in the weak regime of warm inflation.

We conclude this section by finding the lower limit on the friction term which is required for warm
inflation. We shall express this limit in terms of $Q=\Gamma/3H$. Re-write the scalar amplitude Eq.
(\ref{ps}) as
\begin{equation}
{\cal P}_S\approx{T^4\over u^2}{H^3\over T^3}(1+Q)^{1/2}
\end{equation}
The first factor can be replaced using the slow-roll equation Eq. (\ref{sr2}) and the potential
(\ref{corpot}), and we obtain
\begin{equation}
{T\over H}\approx\left({45\over 2\pi^2g_*}\right)^{1/3}(1+Q)^{1/6}
\left( {Q\over {\cal P}_S}\right)^{1/3}.
\end{equation}
The condition for warm inflation $T>H$ is therefore
\begin{equation}
Q>g_*{\cal P}_S.
\end{equation}
Cosmic microwave background observations give a scalar power spectrum of $1\times 10^{-10}$ on large
scales, therefore very small amounts of dissipation can result in warm inflation.

\section{conclusion}

We shall recapitulate the main points of this paper. There are conditions on six of the parameters
defined in Sect. \ref{be} for the possibility of a stable period of warm inflation in the early
universe:
\begin{itemize}
\item The parameter $Q$ which measures the strength of the friction term must satisfy
\begin{equation}
Q>g_*{\cal P}_S.
\end{equation}
where $g_*$ is the effective particle number and ${\cal P}_S$ is the scalar perturbation power
spectrum on large scales.
\item The parameters which describe the inflaton dependence of the effective potential and friction
term satisfy
\begin{equation}
\epsilon<<1+Q,\qquad |\eta|<<1+Q,\qquad |\beta|<<1+Q.
\end{equation}
\item The temperature dependence of the potential and the friction term is restricted by
\begin{equation}
|b|<<{Q\over 1+Q},\qquad |c|<4.
\end{equation}
The condition on $b$ implies that warm inflation is only possible when a mechanism, such as
supersymmetry, reduces the size of the thermal corrections to the potential.
\end{itemize}
Models of elementary particles exist in which these conditions can be satisfied. The most convincing
of these models use a combination of supersymmetry and a two-stage decay process, where there are
no direct coupling between the inflaton and the radiation and all thermal effects are supressed by
factors of $T/\Lambda_S$, where $\Lambda_S$ is the supersymmtry breaking scale 
\cite{berera02,BasteroGil:2006vr,BuenoSanchez:2008nc}.

When the conditions listed above are satisfied, then the solutions to the equations of motion
approach a slow-roll approximation during inflation. As a result, large scale density perturbations
have only one degree of freedom, which we identify as the curvature perturbation. (Entropy
perturbations can only be introduced by adding extra degrees of freedom to the system.)

\acknowledgements
Chun Xiong was supported by an O.R.S. scholarship and by the School of Mathematics and Statistics,
Newcastle University.

%%%%%%%%%%%%%%%%%%%%%%%%%%%%%%%%%%%%%%%%%%%%%
\bibliography{paper.bib}

\end{document}